\documentstyle[sprocl,epsf]{article}

\def\Journal#1#2#3#4{{#1} {\bf #2}, #3 (#4)}

\def\NPB{{\em Nucl. Phys.} B}
\def\PLB{{\em Phys. Lett.}  B}
\def\PRL{\em Phys. Rev. Lett.}

\newcommand{\nc}{\newcommand}
\nc{\beq}{\begin{equation}}
\nc{\eeq}{\end{equation}}
\nc{\beqa}{\begin{eqnarray}}
\nc{\eeqa}{\end{eqnarray}}
\nc{\lsim}{\mbox{\raisebox{-.6ex}{~$\stackrel{<}{\sim}$~}}}
\nc{\gsim}{\mbox{\raisebox{-.6ex}{~$\stackrel{>}{\sim}$~}}}
\def\dr{{\rm d}}
\def\nue{{\nu_e}}

\def\nut{{\nu_\tau}}
\def\bnue{{{\bar \nu}_e}}

\def\nutm{{\nu_{\tau -}}}
\def\nutp{{\nu_{\tau +}}}

\def\mnt{{m_{\nu_\tau}}}
\def\tg{{T_\gamma}}

\def\T100{{T^{100}_{\rm QCD}}}

\def\nn{{\nonumber}}
\def\dN{{\Delta N}}
\def\dNu{{\dN_\nu}}
\begin{document}

\title{PRECISE NUCLEOSYNTHESIS LIMITS ON NEUTRINO MASSES}

\author{Kimmo Kainulainen}

\address{CERN, CH-1211, Geneva 23, Switzerland}

%%%%%%%%%%%%%%%%%%%%%%%%%%%%%%%%%%%%%%%%%%%%%%%%%%%%%%%%%%%%%%
% You may repeat \author \address as often as necessary      %
%%%%%%%%%%%%%%%%%%%%%%%%%%%%%%%%%%%%%%%%%%%%%%%%%%%%%%%%%%%%%%

\maketitle\abstracts{
A computation of nucleosynthesis bounds on the masses of long-lived 
Dirac and Majorana neutrinos\cite{FKO} is reviewed.  In particular
an explicit treatment of the ``differential heating'' of the $\nue$ 
and $\bnue$ ensembles due to the residual out-of-equilibrium 
annihilations of decoupled heavy neutrinos is included.  The effect 
is found to be considerably weaker than originally reported by
Dolgov et al. \cite{DPV}. For example, the bounds for a Dirac tau 
neutrino are $\mnt < 0.37$ MeV or $\mnt > 25$ MeV (for $\dNu > 1$), 
whereas the present laboratory bound is $\mnt < 23.1$ MeV 
\cite{labnew}. }
  
\section{Introduction}

Nucleosynthesis considerations have proved to be an effective tool
in finding limits on particle properties such as masses, couplings,
lifetimes, neutrino mixing parameters and so on.  The constraining
power of nucleosynthesis sensitively depends on the constraints on 
primordial light element abundances that must be inferred from 
the observational evidence.  This is the difficult part of NS 
considerations (see K.A.\ Olive, these proceedings).  Nevertheless, 
as far as NS bounds on new physics are concerned, these details 
can be summarized by a single parameter: the number of equivalent 
neutrino degrees of freedom $\dNu$. 
The theoretical prediction for the helium abundance on the other 
hand, with or without new physics, is relatively straightforward
and can usually be done very accurately.  Parametrizing the 
deviation of the prediction from the standard result in units 
of $\dNu$, one obtains a mapping of the nucleosynthesis bound  
on the space of parameters of the model. This is the situation 
in particular for long-lived massive  neutrinos, and this is what 
I mean with ``precise' nucleosynthesis bounds.

Massive annihilating particles affect the helium abundance indirectly
by altering the expansion rate, or by directly altering the rate of 
reactions holding neutrons and protons in equilibrium:
\beq
n + \nu_e \leftrightarrow  p + e^- \qquad \qquad
n + e^+  \leftrightarrow  p + \bar \nu_e, 
\label{npequ}
\eeq
during the time $T_\gamma \simeq 0.7$ MeV when the $n/p$ ratio is 
freezing out. Also, the changes in the expansion rate alter the
time available for the free neutron decay $n \rightarrow
p + e + \bar \nue$.

For example, additional mass density speeds up the expansion rate
relative to reactions (\ref{npequ}), which causes the $n/p$ ratio 
to freeze out earlier, leaving
behind more neutrons and hence eventually more helium than in the 
reference case of 3 massless neutrinos.
Secondly, since ${\cal O}(few)$ MeV neutrinos freeze out at rather 
low temperatures, their annihilations to $\nue\bar \nue$ final states 
at temperatures below the {\em chemical} freeze-out temperature of 
$\nue$'s ($T_{\rm chem} \simeq 2.3$ MeV) can produce some exess in
the $\nue$ and  $\bnue$ number densities (``bulk heating''), leading 
to an increase of the overall rate of eqs.\ (\ref{npequ}). 
As a result equilibrium is maintained longer, leading to {\em 
less} neutrons and eventually less helium being produced.  In section 
2 I will review a computation that accurately accounts for these two 
effects.

Moreover, the residual annihilations of already decoupled heavy
neutrinos below $T_{\rm kin} \simeq 1$ MeV, when $\nue$'s fall from 
{\em kinetic} equilibrium, can cause a deviation from equilibrium
in the tail of the $\nue$ and $\bnue$ spectra\cite{DPV}. 
Although small in amplitude, the effect of these neutrinos is boosted 
by the quadratic dependence on energy of reactions (\ref{npequ});  
the detailed balance argument relating the {\em equilibrium} conversion 
rates $n \rightarrow p$ and $p \rightarrow n$ does not apply, and 
simply because there are more protons than neutrons around, the
presence of this distortion has the tendency of {\em increasing} 
the relative number of neutrons and, eventually, the final 
helium abundance. I will compute the contribution to $\dNu$ from 
this ``differential heating'' in section 3.  All effects will be 
included in the final results, to be presented in section 4. 

\section{Elements of the Computation}

The relevant momentum-dependent Bolzmann equations for the scalar
phase-space distribution functions have the form:
\beq
E_i(\partial_t - pH\partial_p)f_i(p,t) = C_{{\rm E},i}(p,t)
+ C_{{\rm I},i}(p,t),
\label{bolzmann1}
\eeq
where $E_i = (p^2 + m_i^2)^{1/2}$ and
$H = (8\pi\rho/3 M_{\rm Pl}^2)^{1/2}$ is
the Hubble expansion rate (with  $M_{\rm Pl}$ the Planck mass 
and $\rho$ the total energy density). The index $i$ runs over all 
particle species in the plasma; each distribution function of each 
species has an equation like (\ref{bolzmann1}), and all of them are 
coupled through the elastic and inelastic collision terms
$C_{\rm E}(p,t)$ and $C_{\rm I}(p,t)$.

Since the elastic scattering rates are much higher than 
the annihilation rates, neutrinos freeze out from the chemical 
equilibrium remaining in good kinetic contact with the other 
particles in the plasma. Their spectra can then be described
to a very good accuracy by the pseudo-chemical potentials 
$z_i(T)$, i.e.\ ($E_i \equiv \sqrt{p^2+m^2_i}$)
\beq
f(p,z_i) \equiv (e^{E_i/T_\nu + z_i}+1)^{-1}; \qquad
T_\nu \equiv \left( \frac{4+2h_e(\tg )}{11}\right)^{1/3}
 \:\tg,
\label{fansaz}
\eeq
where $h_e(T)$ is related to the entropy stored in the electrons
and positrons: $s_e \equiv 2\pi^2h_eT^3/45$. 
With the introduction of pseudo-chemical potentials the Bolzmann 
equations (\ref{bolzmann1}) can be integrated over the momenta. One
then has a set of coupled ordinary differential equations for the 
pseudo-chemical potentials of each neutrino species and the photon 
temperature $\tg$\cite{FKO}.

\begin{figure}
\vskip -0.9truecm
\hspace{10truecm}
\epsfysize=11truecm\epsfbox{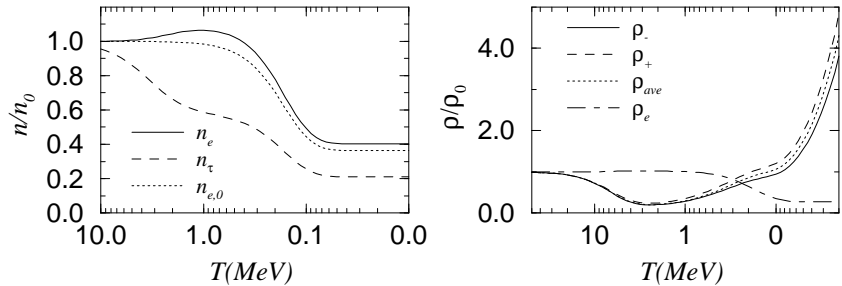}
\vspace{-6.3truecm}
\caption{ 1a) Electron and tau neutrino number densities with a
Majorana $\nut$ with $\mnt=5$ MeV, normalized to equilibrium density 
$n_0 \equiv (3\zeta(3)/2\pi^2)\tg^3$.  The dashed line for comparison 
shows the unperturbed electron neutrino temperature.
1b) Energy densities of $\nutm$, $\nutp$ and $\nue$ with a Dirac $\nut$
with $\mnt = 20$ MeV, normalized to $\rho_0 = 7\pi^2\tg^4/240$. Also shown 
is the energy density corresponding to $\nut$ in the helicity averaged
approximation.}
\end{figure}

In fig.\ 1a a particular solution of the equation network with a 
mass $m^M_{\nut} = 5$ MeV is shown. The importance of tracking the
$\nue$ density is clearly visible: the number density of $\nue$'s remains 
close to the equilibrium value until $T \simeq 2.3$ MeV, after which it 
freezes from chemical equilibrium.  Since annihilations of $\nut$'s 
are still occurring at $T\lsim 2$ MeV, a slight heating of the 
electron neutrino distributions results.  In the units $\dNu$, the 
corresponding effect is\cite{EKT} $\delta \dNu \simeq -3.6 
\delta n_{\nu_e}$.  In this example $\delta n_\nue \simeq 0.1$ 
so that one expects to get $\delta \dNu \sim -0.4$, in good agreement 
with the full nucleosynthesis computation\cite{FKO}.

In the Dirac case there is the additional complication of different
helicity states having different interaction strengths at high energies:
positive helicity states will freeze out earlier with larger relic
density (and the negative helicity states later with smaller) than 
the hypotetical Dirac neutrino with helicity-averaged interaction 
strength. One sees this effect in fig.\ 1b, which shows the energy
density stored into different helicities. These deviations almost exactly 
cancel each other in the sum, however, so that the total number 
density, and hence the induced effect on helium synthesis, is very 
close to the results found using the helicity-averaged approach.  
Figure 1b also shows how the relative energy density stored into
tau neutrinos sharply increases at small $T$.

\section{Differential Heating}

Qualitatively the physics was explained in the introduction; for more 
details see Dolgov et al.\ \cite{DPV}.
Since one is considering a very small amplitude effect in the 
electron neutrino distributions, one can to a good accuracy use
the linearized Bolzmann equation for the deviation.  Moreover, because
the effect is dominated by large energies, it is an exellent 
approximation to replace the Fermi--Dirac distributions by Bolzmann 
distributions.  Then noting that the l.h.s.\ of eq.\ (\ref{bolzmann1}) 
annihilates the equilibrium distribution, and trading the variables 
$(t,p)$ for $x_\nu \equiv \mnt/T_\nu$ and $y\equiv p/\tg$, 
one gets the generic equation for the deviation:
\beq
H x_\nu \frac{\partial }{\partial x_\nu} \delta f(x,y) = 
C_{\rm ann}(x,y) - C_{\rm el}(x,y) \delta f(x,y),
\label{diffbe}
\eeq
where $x \equiv \mnt/T$.  The variables $x$ and $x_\nu$ are of course
simply related; using $x_\nu$ allowed succinctly including the effect
of entropy release in eq.\ (\ref{diffbe}). The annihilation term 
$C_{\rm ann}(x,y)$ acts as a source and it is peaked around $y\simeq x$.  
The term $C_{\rm el}(x,y)$ represents the restoring force of the 
elastic scatterings. In magnitude, $C_{\rm el} \gg C_{\rm ann}$.  
Annihilation terms are easily computed, and I find the forms similar 
to the ones by Dolgov et al.\ \cite{DPV}. On the Dirac case, for 
example
\beq
C^D_{\rm ann} \simeq  9.0\times 10^{-3} m^3_{\nut} 
\frac{n(x)^2-n_{\rm eq}(x)^2}{x^4\sqrt{y}} \times 
\exp [-(x-y)^2/y] F_D(x,y),
\label{cann}
\eeq
where $n$'s are the number densities normalized as in 
fig.\ 1 and  $F_D(x,y)$ is a function with the asymptotic value 
of 1 for $y\rightarrow 0, x\rightarrow \infty$ \cite{DPV}. 
In numerical work I used exact functions, valid for any $x$ and 
$y$, but this is not a significant effect. The elastic scattering 
term is given by ($n_{\rm eq}(0) \equiv 1$)
\beq
C_{\rm el}(x,y) \simeq 0.75 \frac{m^5y}{x^5}\left(\frac{T_\nu}{\tg }\right)^4
   \left( 1 + 0.23n_{eq}(x_\nu)
            + 0.54\left(\frac{\tg }{T_\nu}\right)^4n_{eq}(m_e/\tg )\right).
\label{cel}
\eeq
The solution of (\ref{diffbe}) is straightforward:
\beq
\delta f(x,y) = \int_0^x {\rm d} x' A(x') C_{\rm ann}(x',y)\; 
\exp \left[ -\int_{x'}^x {\rm d}x'' A(x'') C_{\rm el}(x'',y) \right],
\label{gensol}
\eeq
where
\beq
A(x) \equiv \frac{1}{x H(x)} \frac{\tg }{T_\nu} 
    \left( 1+\frac{\tg}{3h_I}\frac{\dr h_I}{\dr \tg}\right),
\eeq
and $h_I(\tg ) = 2 + h_e(\tg )$. Equations (\ref{diffbe}) and 
(\ref{gensol}) differ from those of Dolgov et al.\ \cite{DPV} 
in that I included the dilution due  to the entropy release in the 
electron annihilations, as well as the correct energy dependence in the 
expansion rate $H$, both of which tend to weaken the effect. Moreover, 
the elastic scattering term (\ref{cel}) is somewhat larger than that of
Dolgov et al., which also weakens the effect.  
Fig.\ 2a displays a particular solution of 
equation (\ref{gensol}) and final results for the effect of differential 
heating are shown in figure 2b.  The effect is largest for the Dirac 
neutrino at around $\mnt \simeq 10$ MeV.  However, for $\mnt = 25$ MeV 
it is already below 0.25 effective neutrino degrees of freedom, 
i.e.\ a fourth of that found by Dolgov et al.  These results are in very 
good agreement with the full numerical solution of the Bolzmann equation 
by Hannestad and Madsen\cite{HM} in the Majorana case.

\begin{figure}
\vskip -0.9truecm
\hspace{10truecm}
\epsfysize=11truecm\epsfbox{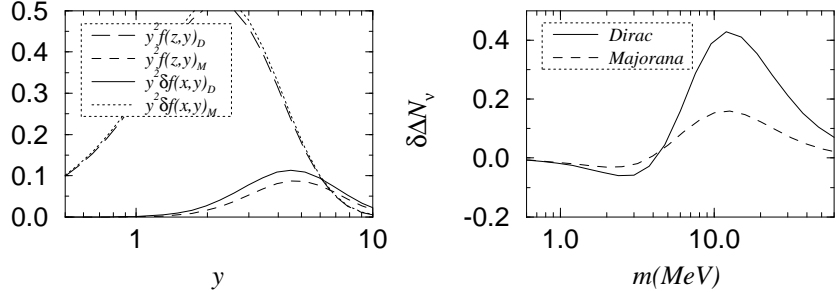}
\vspace{-6.1truecm}
\caption{a) The distributions $y^2\delta f(x,y)$ for $m_\nut = 10$ MeV,
and $T=1$ MeV ($x=10$) are shown with the corresponding bulk-equilibrium 
distributions.  b) The effect on BBN of the differential heating in units 
of equivalent $\nu$ degrees of freedom.}
\end{figure}

\section{Results}

The mass bounds can be expressed in terms of fit functions
of $\dNu$. Including all the effects to the electron neutrino
distributions discussed above, we find that the Majorana
masses are bounded by
\beqa
m^M_\nu &<&  \sqrt{x}(0.350 + 0.047\sqrt{x} + 0.591x)
              \; \theta (0.15-x) \nn \\
        &+&  (0.079 + 0.576x + 0.598x^2 - 0.514x^3 + 0.211x^4)
              \; \theta (x-0.15) \nn \\\
m^M_\nu &>&  68.59 - 63.83x + 49.18x^2 - 33.09x^3 + 13.18x^4 - 2.17x^5,
\label{majbnd}
\eeqa
where the units are  MeV and I used $x\equiv \dNu$.
One sees that opening up a window for a stable tau neutrino 
below the laboratory bound of 23.1 MeV \cite{labnew} would require 
relaxing the nucleosynthesis bound to $\dNu > 1.44$. Then the NS 
bound of $\dNu < 1$, together with the above laboratory bound rules 
out a long-lived Majorana tau neutrino with $m^M_\nut > 0.95$ MeV.

In the Dirac case the upper bound on the disallowed region leads to 
the constraint
\beq
m^D_\nut > 41.88 - 28.47x + 20.44x^2 - 13.60x^3 + 5.49x^4 - 0.90x^5,
\label{dirlbnd}
\eeq
where the units are again MeV. The effect of differential 
heating was for $\dNu >1$ to increase the bound from 22 MeV to
$\mnt > 25$ MeV, closing the window below the experimental bound
of $\mnt <23.1$ MeV at 95\% CL.

The computation of the lower bound on the disallowed region
is entirely different from that in the preceding cases and completely 
unchanged by the differential heating effects.  I will only quote 
the results from Fields et al.: for the bound $\dNu >1$ they give
$m_{\nu_\mu} \lsim 0.31$ MeV and $m_{\nu_\tau}\lsim 0.37$ MeV,
with $T_{QCD}=100$ MeV.  These constraints are conservative in the
sense that they would be somewhat stricter if the QCD transition 
temperature was chosen higher\cite{FKO}.

In conclusion, even with the weak constraint of $\dNu < 1$,
nucleosynthesis bound is strong enough to exclude a long-lived 
($\tau \gsim 100$ sec) Majorana tau neutrino with $\mnt > 0.95$ MeV 
and a Dirac tau neutrino with $\mnt > 0.37$ MeV.

\section*{Acknowledgements}

I wish to thank Sacha Dolgov, Sten Hannestad and Jes Madsen for  
useful discussions during the Neutrino 96 conference in Helsinki.

\section*{References}
%
% ---------------------------------------------------------------------
% REFERENCES
% ---------------------------------------------------------------------
%

\end{document}